\documentclass{webofc}
\usepackage[varg]{txfonts}   
\begin{document}
\title{Computing one-loop radiative corrections in $\tau \to \pi (K) \nu [\gamma]$ and testing new physics}
%
%

\author{\firstname{Ignasi} \lastname{Rosell}\inst{1} \fnsep
\thanks{\email{rosell@uchceu.es}} 
}

\institute{Departamento de Matem\'aticas, F\'\i sica y Ciencias Tecnol\' ogicas, Universidad Cardenal Herrera-CEU, CEU Universities, 46115 Alfara del Patriarca, Val\`encia, Spain 
          }

\abstract{%
  Without a doubt the ratios $R_{\tau / P}\equiv \Gamma \left( \tau \to P \nu_{\tau}[\gamma] \right) / \Gamma \left(P\to \mu \nu_{\mu}[\gamma]\right)$ ($P=\pi,K$) give a convenient scenario to test the lepton universality, the CKM unitarity and to search for non-standard interactions in $\tau$ decays. Moreover, the previous theoretical estimation of these observables is more than twenty-five years old and some assumptions of that estimation are unrealistic. Therefore, an update of $R_{\tau / P}$ was timely. The one-loop radiative corrections have been computed by considering an expansion of Chiral Perturbation Theory including the lightest spin-one resonances and respecting the short-distance behavior dictated by QCD. We have reported $\delta R_{\tau/\pi}=(0.18\pm 0.57 )\%$  and $\delta R_{\tau/K}=(0.97\pm 0.58 )\%$, where the uncertainties have been induced fundamentally by the estimation of the counterterms. We have tested the lepton universality, the CKM unitarity and have searched for new physics in $\tau$ decays. As a by-product, we have also determined the theoretical radiative corrections to the $\tau \to P \nu_{\tau}[\gamma]$ decay rates, $\delta_{\tau \pi} = -(0.24 \pm 0.56) \%$ and $\delta_{\tau K} = -(0.15 \pm 0.57) \%$.

}
\maketitle
\section{Introduction}
\label{intro}

Although within the Standard Model the three lepton families are expected to couple to the electroweak gauge bosons with the same intensity $g_{\ell}$ ($\ell=e,\, \mu,\, \tau)$, the so-called lepton universality (LU), and this has been tested experimentally and most of the results have been compatible with LU, a few anomalies observed in semileptonic $B$ meson decays~\cite{Albrecht:2021tul} seem to indicate a small deviation from LU. Therefore, an interesting task is the study of all the different observables testing LU, including low-energy precision probes using pion, kaons and tau leptons, which currently provide the most precise tests of LU~\cite{Bryman:2021teu}. Following this idea, we have tested LU between the second and third families through the ratio ($P=\pi,K$)~\cite{Marciano, DF,nosotros}
\begin{equation}
R_{\tau / P}\equiv \frac{ \Gamma \left( \tau \to P \nu_{\tau}[\gamma] \right)}{\Gamma \left(P\to \mu \nu_{\mu}[\gamma]\right)}= \left\vert \frac{g_\tau}{g_\mu} \right\vert^2_P R_{\tau/P}^{0}(1+\delta R_{\tau / P}), \label{MainDef}
\end{equation}
where $g_\mu=g_\tau$ according to LU, $\delta R_{\tau/P}$ denotes the radiative corrections, and $R_{\tau/P}^{(0)}$ is the tree-level contribution given by
\begin{equation}
R_{\tau/P}^{(0)}=\frac{1}{2}\frac{M_\tau^3}{m_\mu^2 m_P}\frac{(1-m_P^2/M_\tau^2)^2}{(1-m_\mu^2/m_P^2)^2}, \label{LO}
\end{equation}
which is free from quark mixing angles and hadronic couplings.

In Ref.~\cite{DF}, more than twenty-five years ago, the first complete estimation for $\delta R_{\tau/P}$ was given: $\delta R_{\tau/\pi}=(0.16\pm0.14)\%$ and $\delta R_{\tau/K}=(0.90\pm0.22)\%$. However, there were important reasons to analysis this observable again: the hadronic form factors of Ref.~\cite{DF}  are different for virtual- and real-photon contributions, do not follow the correct QCD high-energy behavior, violate analyticity, unitarity, and the chiral limit at leading non-trivial orders; moreover, a cutoff is implemented in order to regulate loop integrals, splitting unphysically short- and long-distance contributions. Besides, errors reported in Ref.~\cite{DF} are not realistic, since they are of the order of an expected $\mathcal{O}(e^2p^2)$ Chiral Perturbation Theory (ChPT) calculation, whose only model-dependence is usually the estimation of the counterterms, but it does not allow the inclusion of the $\tau$ lepton.

Depending on the process at hand, different values of $\left|g_\tau/g_\mu\right|$ can be found in the literature:
\begin{enumerate}
\item $\Gamma(\tau \to P \nu_\tau  [\gamma]) / \Gamma(P \to\mu \nu_\mu[\gamma])$ ($P=\pi,K$). Considering the values of $\delta R_{\tau/P}$ given in Ref.~\cite{DF}, the 2018 HFLAV analysis~\cite{Amhis:2019ckw} reported $\left|g_\tau/g_\mu\right|_\pi=0.9958\pm0.0026$ and $\left|g_\tau/g_\mu\right|_K=0.9879\pm0.0063$, at $1.6\sigma$ and $1.9\sigma$ of LU.
\item $\Gamma(\tau \to e \bar{\nu}_e \nu_\tau[\gamma])/\Gamma(\mu \to e \bar{\nu}_e \nu_\mu[\gamma])$. This purely leptonic extraction gives $\left|g_\tau/g_\mu\right|=1.0010\pm0.0014$~\cite{Amhis:2019ckw},  at $0.7\sigma$ of LU.
\item $\Gamma(W \to\tau \nu_\tau)/\Gamma(W \to\mu \nu_\mu)$. The weighted average of the $W$-boson decay determinations yields $\left|g_\tau/g_\mu\right|=0.995\pm 0.006$~\cite{Aad:2020ayz, CMS:2021qxj}, at $0.8\sigma$ of LU.
\end{enumerate}

Thus, a new estimation of $\delta R_{\tau/P}$ was interesting to solve these discrepancies, being the improvement of the theoretical Ans\"atze an important ingredient.

In Ref.~\cite{nosotros} we have presented a new next-to-leading calculation of $\delta R_{\tau/P}$ considering a large-$N_C$ effective approach including the lightest resonances~\cite{RChT} and overcoming all the aforementioned theoretical difficulties. Note that whereas $P$ decays can be analyzed unambiguously by using Chiral Perturbation Theory (the Standard Model), being the estimation of the local counterterms the only model dependence, $\tau$ decays must be studied considering a model-dependent effective approach encoding the hadronization of the QCD currents: this is the reason why we have considered in Ref.~\cite{nosotros} the large-$N_C$ approach quoted previously~\cite{RChT}. 

The new calculation of $\delta R_{\tau/P}$ has been used not only to analyze the LU in $\left|g_\tau/g_\mu\right|$, but also to report the theoretical radiative corrections in individual $\tau \to P \nu_{\tau}[\gamma]$ decay rates, to test the CKM unitarity via $\left|V_{us}/V_{ud}\right|$ in $\Gamma(\tau \to K \nu_\tau[\gamma])/\Gamma(\tau \to \pi \nu_\tau[\gamma])$ or through $\left|V_{us}\right|$ in $\Gamma(\tau \to K \nu_\tau[\gamma])$ and to search for non-standard interactions on these decays. It is interesting to stress that our results have been used in the recent 2021 HFLAV analysis~\cite{HFLAV:2022pwe} instead of Ref.~\cite{DF}.

\section{The calculation}

Despite their different approaches, $P \to\mu \nu_\mu[\gamma]$ and $\tau \to P \nu_\tau  [\gamma]$ decay rates can be organized similarly~\cite{Marciano,nosotros,CR}:
\begin{align}
\Gamma_{P_{\mu 2 [\gamma]}} =\,& 
\Gamma^{(0)}_{P_{\mu 2 }} \!
\, S_{\rm EW} \Bigg\{ 1 + \frac{\alpha}{\pi}  \,  F( m_\mu^2/m_P^2)  \Bigg\}
\Bigg\{  1 - \frac{\alpha}{\pi}   \Bigg[ \frac{3}{2}  \log \frac{m_\rho}{m_P}\nonumber\\
&+   c_1^{(P)}  + \frac{m_\mu^2}{m_\rho^2}   \bigg(c_2^{(P)}  \, \log \frac{m_\rho^2}{m_\mu^2}   +  c_3^{(P)} 
+ c_4^{(P)} (m_\mu/m_P) \bigg)  -  \frac{m_P^2}{m_\rho^2} \,  \tilde{c}_{2}^{(P)}  \, \log \frac{m_\rho^2}{m_\mu^2}  \Bigg] \Bigg\} \,,
\label{eq:indrate} \\
\Gamma_{\tau_{P2[\gamma]}}=\,&
\Gamma^{(0)}_{\tau_{P2}}
\, S_{\rm EW}
\Bigg\{ 1 + \frac{\alpha}{\pi}  \,  G (m_P^2/M_\tau^2)  \Bigg\}
\Bigg\{  1 - \frac{3\alpha}{2\pi}  \log \frac{m_\rho}{M_\tau} +  \delta_{\tau P}\big|_{\mathrm{rSD}} +  \delta_{\tau P}\big|_{\mathrm{vSD}}\Bigg\} \,,
\label{eq:indratetau}
\end{align}
where $S_{\rm EW}=1.0232\simeq  1 + \frac{2 \, \alpha}{\pi}  \log \frac{m_Z}{m_\rho}$ corresponds to the (universal) leading short-distance electroweak correction~\cite{Marciano} (canceling in the ratio  $R_{\tau/P}$), the first bracketed terms are the universal long-distance or point-like correction~\cite{DF,nosotros,Kinoshita:1959ha}, the second bracketed terms include the structure-dependent (SD) contributions and $\Gamma^{(0)}_{P_{\mu 2 }}$ and $\Gamma^{(0)}_{\tau_{P2}}$ are the decay rate at leading order ($F_\pi\sim92$ MeV),
\begin{equation}
\Gamma^{(0)}_{P_{\mu 2}} =  \frac{G_F^2 |V_{uD}|^2  F_P^2 }{4 \pi} \,  m_P  \, m_\mu^2  \, \left(1 - \frac{m_\mu^2}{m_P^2} \right)^2 \,,  \qquad
\Gamma^{(0)}_{\tau_{P2}}=\frac{G_F^2 |V_{uD}|^2F_P^2 }{8\pi} M_\tau^3\left(1-\frac{m_P^2}{M_\tau^2}\right)^2, 
\label{eq:indratetauLO}
\end{equation}
being $D=d,s$ for $P=\pi,K$, respectively.  Whereas for structure-dependent contributions in $P_{\mu 2}$ the notation of Ref.~\cite{Marciano} has been used and the numerical values for $c_n^{(P)}$ in Ref.~\cite{CR} have been considered, we have separated the structure-dependent contributions in $\tau_{P2}$ into real-photon (rSD) and virtual-photon (vSD) corrections: 
we have taken rSD corrections from Ref.~\cite{Guo:2010dv} and have performed a new calculation for vSD ones. 

\begin{table*}[t!!!!]
\begin{center}
\renewcommand{\arraystretch}{1.5}
\begin{tabular}{|c|c|c|c|}
\hline
  Contribution & $\delta R_{\tau/\pi}$   & $\delta R_{\tau/K}$ &  Ref.  \\[5pt]
\hline \hline 
SI &  $+1.05\%$& $+1.67\%$ &\cite{DF} \\
rSD  &$+0.15\%$   &$+(0.18\pm 0.05)\%$ & \cite{CR,Guo:2010dv} \\
vSD & $-(1.02\pm 0.57 )\%$& $-(0.88\pm 0.58)\%$ & \cite{nosotros,CR} \\
\hline \hline
Total & $+(0.18\pm 0.57 )\%$ & $+(0.97\pm 0.58 )\%$ & \cite{nosotros}
\\
\hline
\end{tabular}
\end{center}
\caption{Numerical values of the different contributions to $\delta R_{\tau/P}$: Structure Independent (SI), real-photon Structure Dependent (rSD) and virtual-photon Structure Dependent (vSD). Uncertainties are not given if they are negligible for the level of accuracy of this analysis, that is, lower than $0.01\%$.}
\label{tab:tab1}
\end{table*}

The technicalities of the calculation are given in Ref.~\cite{nosotros} and the different photonic contributions are showed in Table~\ref{tab:tab1} . The final result reads~\cite{nosotros}
\begin{equation} 
\delta R_{\tau/\pi} \,=\, (0.18\pm 0.57 )\% \,, \qquad \qquad
\delta R_{\tau/K}\,=\, (0.97\pm 0.58 )\%\,. \label{finalresult}
\end{equation} 
The most relevant source of uncertainty comes from the estimation of the counterterms appearing in vSD contributions of $\tau \to P \nu_\tau$ ($\pm 0.57\%$ and $\pm 0.58\%$ for the pion and kaon case, respectively, see Table~\ref{tab:tab1}). This uncertainty has been estimated by considering the running of the counterterms between $0.5\,$and $1.0\,$GeV ($\pm 0.52\%$, in a similar way to Ref.~\cite{Marciano}) and considering the effect of assuming a less general resonance effective Lagrangian.\footnote{Let us stress the conservative estimation followed here. Due to the different scales, counterterms in effective approaches with resonances are expected to be lower than in ChPT, without resonances. Neverthesless, with our estimation the counterterms affecting the vSD corrections in $P_{\mu 2}$ and $\tau_{P2}$ are assumed to be of similar size.}

Although the final result reported in (\ref{finalresult}) is compatible with Ref.~\cite{DF}, $\delta R_{\tau/\pi}=(0.16\pm0.14)\%$ and $\delta R_{\tau/K}=(0.90\pm0.22)\%$, the comparison needs to be done carefully. Firstly, in our view uncertainties were underestimated in Ref.~\cite{DF}, for they have approximately the expected size in a purely Chiral Perturbation Theory computation, a much more model-independent scenario. And secondly, we would like to stress again the inconsistencies in Ref.~\cite{DF}: the hadronic form factors are different for real- and virtual-photon corrections, do not satisfy the asymptotic behavior dictated by QCD, violate analyticity, unitarity and the chiral limit at leading non-trivial orders, and a cutoff is used, separating for no reason short- and long-distance contributions. Moreover, as it is explained in Appendix B of Ref.~\cite{nosotros}, despite the fact that central values of (\ref{finalresult}) are very similar to the ones reported in Ref.~\cite{DF}, central values of the different SD corrections are very different within both analysis, as it can be seen in Table~\ref{tab:tab2}:
\begin{enumerate}
\item Virtual corrections by Ref.~\cite{DF} are cutoff-dependent, since, as it has been spotlighted previously, long- and short-distance photonic contributions are separated unphysically: figures with an asterisk are cutoff-dependent.
\item The reported error in the radiative correction of Ref.~\cite{DF} arises from uncertainties in hadronic parameters of SD contributions and from variations in $\mu_{\rm cut}$.
\item For the SI contribution in Ref.~\cite{DF} we have added to the result obtained in the point-like approximation ($1.05\%$), the term coming from cutting off the loops at $\mu_{\rm cut}$ ($-0.21\%$).
\item Different contributions of $\delta R_{\tau/K}$ are not provided in Ref.~\cite{DF}, preventing a comparison.
\end{enumerate}
Consequently, the agreement between the central values seems to be only a coincidence.

\begin{table}[t!!!!]
\begin{center}
\begin{tabular}{|c|c|c|}
\hline
  Contribution & Ref.~\cite{DF} [$\mu_{\rm cut}=1.5~$GeV]& Ref.~\cite{nosotros}\\
\hline \hline 
SI & $+0.84\%^*$ & $+1.05\%$  \\
rSD & $+0.05\%$ & $+0.15\%$ \\
vSD & $-0.49\%^*$ & $-(1.02\pm 0.57)\%$ \\
short-distance & $-0.25\%^*$ & 0 \\
\hline \hline
Total & $+(0.16\pm 0.14)\%^*$ & $+(0.18\pm 0.57)\%$ \\
\hline
\end{tabular}
\end{center}
\caption{Comparison of photonic contributions to $\delta R_{\tau/\pi}$~\cite{nosotros}. The asterisk indicates that the figure at hand depends on the cutoff, which has been established at $\mu_{\rm cut}=1.5$ GeV \cite{DF}.} \label{comparison}
\label{tab:tab2}
\end{table}

\section{Applications}

As it has been explained in the introduction, there are different interesting applications of our result~\cite{nosotros}:
\begin{enumerate}
\item Radiative corrections in $\tau \to P \nu_\tau[\gamma]$ decay rates. We have used our results to estimate the radiative corrections in the individual $\tau_{P2[\gamma]}$ decays, $\Gamma_{\tau_{P2[\gamma]}} =  \Gamma^{(0)}_{\tau_{P2}} S_{\rm EW}\left( 1 +\delta_{\tau P} \right)$, where $\delta_{\tau P}$ includes all SI and SD radiative corrections of (\ref{eq:indratetau}) and we have reported~\cite{nosotros}:
\begin{equation} 
\delta_{\tau \pi} \,=\, -( 0.24 \pm 0.56 ) \%\, , \qquad \qquad
\delta_{\tau K} \,=\, -(0.15 \pm 0.57) \%\, . \label{delta}
\end{equation}
\item Lepton universality test. Considering (\ref{MainDef}) and (\ref{finalresult}), the LU can be tested~\cite{nosotros},
\begin{align} 
&\left|\frac{g_\tau}{g_\mu}\right|_\pi =0.9964 \pm 0.0028_{\mathrm{th}}\pm 0.0025_{\mathrm{exp}}  = 0.9964\pm 0.0038\,, \nonumber \\
&\left|\frac{g_\tau}{g_\mu}\right|_K=0.9857\pm  0.0028_{\mathrm{th}}\pm 0.0072_{\mathrm{exp}}  =0.9857\pm 0.0078\,, \label{finalLU}
\end{align}
at $0.9\sigma$ and $1.8\sigma$ of LU ($g_\tau=g_\mu$), respectively. Note that while experimental and theoretical uncertainties are of similar size in the pion case, experimental ones dominate in the kaon case. We can compare these results with the 2018 HFLAV analysis~\cite{Amhis:2019ckw}, $\left|g_\tau/g_\mu\right|_\pi=0.9958\pm0.0026$ and $\left|g_\tau/g_\mu\right|_K=0.9879\pm0.0063$ (at $1.6\sigma$ and $1.9\sigma$ of LU), where $\delta R_{\tau/P}$ was taken from Ref.~\cite{DF}. In the recent HFLAV analysis of Ref.~\cite{HFLAV:2022pwe}, where our results of (\ref{finalresult}) have been considered, the agreement is larger: $\left|g_\tau/g_\mu\right|_\pi=0.9959\pm0.0038$ and $\left|g_\tau/g_\mu\right|_K=0.9855\pm0.0075$ (at $1.1\sigma$ and $1.9\sigma$ of LU).
\item CKM unitarity test via $\left|V_{us}/V_{ud}\right|$. One can use the ratio
\begin{equation}\label{eq:Vusdet}
\frac{\Gamma(\tau\to K\nu_\tau[\gamma])}{\Gamma(\tau\to \pi\nu_\tau[\gamma])}=\frac{|V_{us}|^2F_K^2}{|V_{ud}|^2F_\pi^2}\frac{(1\!-\!m_{K}^2/M_\tau^2)^2}{(1\!-\!m_{\pi}^2/M_\tau^2)^2}\left(1\!+\!\delta\right)\!,
\end{equation}
in order to extract $\left|V_{us}/V_{ud}\right|$. Considering that the radiative correction in (\ref{eq:Vusdet}) can be extracted from (\ref{delta}), $\delta = \delta_{\tau K}- \delta_{\tau \pi}= (0.10\pm0.80)\%$, we have found~\cite{nosotros}:
\begin{align} 
& \bigg|\frac{V_{us}}{V_{ud}}\bigg|=0.2288 \pm 0.0010_{\mathrm{th}} \pm 0.0017_{\mathrm{exp}}  = 0.2288\pm0.0020\,, \label{ourVusVud}
\end{align}
at $2.1\sigma$ of unitarity, being dominant the experimental uncertainties. Despite the fact that this analysis was absent in the 2018 HFLAV analysis~\cite{Amhis:2019ckw}, it has been included in the 2021 version~\cite{HFLAV:2022pwe}, where our result of (\ref{finalresult}) has been considered and it has been reported $|V_{us}/V_{ud}|=0.2289\pm0.0019$, very similar to (\ref{ourVusVud}). Our result is also consistent with Ref.~\cite{Seng:2021nar}, $|V_{us}/V_{ud}|=0.2291\pm0.0009$, obtained in the context of kaon semileptonic decays, where the larger statistics allows lower uncertainties.
\item CKM unitarity test via $\left|V_{us}\right|$. Alternatively, $|V_{us}|$ can be obtained directly from the $\tau \to K \nu_\tau[\gamma]$ decay rate, $\Gamma_{\tau_{K2[\gamma]}} =  \Gamma^{(0)}_{\tau_{K2}} S_{\rm EW}\left( 1 +\delta_{\tau K} \right)$. We have found~\cite{nosotros}: 
\begin{align} \label{ourVus}
 &|V_{us}|=0.2220 \pm  0.0008_{\mathrm{th}}  \pm 0.0016_{\mathrm{exp}}  = 0.2220\pm 0.0018 \,, 
\end{align}
at $2.6\sigma$ of unitarity and again being dominant the experimental uncertainties. This result can be compared with the 2018 HFLAV analysis~\cite{Amhis:2019ckw},  $ |V_{us}|= 0.2234 \pm 0.0015$ and using Ref.~\cite{DF}, or the recent 2021 HFLAV analysis~\cite{HFLAV:2022pwe}, $ |V_{us}|=0.2219\pm 0.0017$ and considering our results. Again (\ref{ourVus}) is compatible with Ref.~\cite{Seng:2021nar}, $|V_{us}|=0.2231\pm0.0006$, obtained using kaon semileptonic decays and, consequently, with a better precision.

\item New physics in $\tau \to P \nu_\tau[\gamma]$ decay rates. We have also used our results to constrain non-SM interactions in $\tau \to P \nu_\tau[\gamma]$ decays: 
\begin{equation}
\Gamma(\tau \to  P\nu_\tau[\gamma]) =  \Gamma^{(0)}_{\tau_{P2}}  \left|\frac{\widetilde{V}_{uD}}{V_{uD}}\right|^2 S_{\rm EW} \left( 1 + \delta_{\tau P}  + 2 \Delta^{\tau P}\right) \,,
\end{equation}
being $D=d,s$ for $P=\pi,K$, respectively. $\Delta^{\tau P}$ contains the leading-order new-physics corrections not present in $\widetilde{V}_{uD}=(1 + \epsilon^e_L + \epsilon^e_R )V_{uD}$, directly incorporated by considering 
$V_{uD}$ from nuclear $\beta$ decays~\cite{EFTtaudecays1,EFTtaudecays2}:
\begin{equation}
\Delta^{\tau P} = \epsilon^\tau_L-\epsilon^e_L-\epsilon^\tau_R-\epsilon^e_R-\frac{m_P^2}{M_\tau(m_u+m_D)}\epsilon^\tau_P \,.
\end{equation}
Considering our estimations of $\delta_{\tau P}$ and $|V_{us}/V_{ud}|$ given in (\ref{delta}) and (\ref{ourVusVud}), respectively, we have estimated~\cite{nosotros}:
 \begin{equation}
\Delta^{\tau \pi} \,=\, -(0.15\pm0.72)\%\,, \qquad \qquad 
\Delta^{\tau K} \,=\,-(0.36\pm1.18)\%\, ,
\end{equation}
to be compared to $\Delta^{\tau \pi} = -( 0.15 \pm 0.67 ) \%$ in Ref.~\cite{EFTtaudecays1}, $\Delta^{\tau \pi} = -( 0.12 \pm 0.68 ) \%$ and $\Delta^{\tau K} = -(0.41 \pm 0.93) \%$ in Ref.~\cite{EFTtaudecays2}, and $\Delta^{\tau \pi} = -( 0.09 \pm 0.73 ) \%$ and $\Delta^{\tau K} = -(0.2 \pm 1.0) \%$ in Ref.~\cite{Cirigliano:2021yto}. All these values are reported at a scale of $\mu=2$ GeV in the $\overline{\mathrm{MS}}$-scheme. Note that all these determinations are consistent with each other and compatible with the SM, $\Delta^{\tau P}=0$.
\end{enumerate}

For all these determinations branching ratios and masses have been obtained from the PDG~\cite{PDG}, $|V_{ud}|=0.97373\pm0.00031$ from Ref.\cite{Hardy:2020qwl}, $S_{\rm EW}=1.0232$ from Ref.~\cite{Marciano} and meson decay constants from the FLAG analysis~\cite{Aoki:2019cca}: $\sqrt{2} F_\pi=(130.2\pm0.8)~$MeV, $\sqrt{2} F_K=(155.7\pm0.3)~$MeV and $F_K/F_\pi=1.1932\pm 0.0019$, .

\vspace{0.5cm}
\begin{acknowledgement}
I wish to thank the organizers of the workshop for the pleasant conference. I am also very grateful to M. A.~Arroyo-Ure\~na, G.~Hern\' andez-Tom\'e, G.~L\'opez-Castro and P.~Roig, since the work presented here has been done with them and also for their helpful comments to prepare these proceedings. This work has been supported in part by the Universidad Cardenal Herrera-CEU [INDI21/15]; by the Generalitat Valenciana [PROMETEU/2021/071] and by the Spanish Government [MCIN/AEI/10.13039/501100011033, Grant No. PID2020-114473GB-I00]. 
\end{acknowledgement}

\end{document}